\PassOptionsToPackage{hyphens}{url}
\documentclass[10pt,conference]{IEEEtran}

\usepackage{cite}
\usepackage{amsmath,amssymb,amsfonts}
\usepackage{booktabs}
\usepackage{graphicx}
\usepackage{float}
\usepackage{xspace}
\usepackage{xcolor}
\usepackage{multirow}
\usepackage{tcolorbox}
\tcbuselibrary{breakable}
\usepackage{listings}
\usepackage[hidelinks]{hyperref}
\usepackage[capitalize,nameinlink]{cleveref}

\lstdefinestyle{prompt}{
  basicstyle=\ttfamily\scriptsize,
  breaklines=true,
  breakatwhitespace=false,
  columns=fullflexible,
  keepspaces=true,
  frame=single,
  framexleftmargin=2pt,
  xleftmargin=2pt,
  showstringspaces=false,
  captionpos=b,
  aboveskip=6pt, belowskip=4pt,
}
\crefname{lstlisting}{Listing}{Listings}
\Crefname{lstlisting}{Listing}{Listings}
\crefname{appendixsec}{Appendix}{Appendices}
\Crefname{appendixsec}{Appendix}{Appendices}

\newcommand{\defi}{DeFi\xspace}

\newcommand{\code}[1]{\texttt{#1}}
\newcommand{\ese}{ESE\xspace}
\newcommand{\topicgpt}{TopicGPT\xspace}
\newcommand{\nquestions}{54,637\xspace}
\newcommand{\ntopicpairs}{72,572\xspace}
\newcommand{\nuniquetopics}{22\xspace}
\newcommand{\datespan}{January 2016 to December 2025\xspace}
\newcommand{\rqueue}[1]{\textbf{RQ\textsubscript{#1}}\xspace}

\newcommand{\hide}[1]{}

\newtcolorbox{roundbox}{
  colback=gray!8,
  colframe=gray!50,
  arc=2mm,
  boxrule=0.4pt,
  left=4pt,right=4pt,top=3pt,bottom=3pt,
  before skip=4pt, after skip=4pt
}

\begin{document}

\title{What Ethereum Developers Ask For, and Whether They Get It: Ten Years of Ethereum Stack Exchange}


\author{\IEEEauthorblockN{Tamer Abdelaziz}
\IEEEauthorblockA{\textit{New York University Abu Dhabi} \\
tamer.m@nyu.edu}}

\maketitle

\begin{abstract}
The Ethereum community learns where smart-contract developers struggle through surveys and interviews, which record what developers say when asked about a chosen theme. Ethereum Stack Exchange (\ese) records what they ask for unprompted, and whether anyone resolves it. We study every question posted to \ese from \datespan, \nquestions in all, and label each with a prompt-based topic pipeline seeded with eight Ethereum-specific topics and open to new ones (\nuniquetopics topics in all). A three-provider committee of language models screens the labels, and manual adjudication of every disputed case puts their accuracy at 87.9 to 93.5\%. Because the corpus is the complete record, no rate carries sampling error. We ask what developers ask for help with, whether the venue still resolves it, and, where it does not, whether the questions changed or their resolution did.

Demand is concentrated on writing, deploying, and debugging contracts; the toolchain is the second-largest topic and growing, while security, consensus, and scaling are thin. Supply has weakened, and unevenly: answers still arrive, closure falls further. Across five estimators of the decade, the share of questions drawing any answer fell by 9.5 to 18.6~percentage points, but among answered questions the share reaching an accepted answer fell by 16.2 to 24.5. Rebuilding the later acceptance rate at the earlier topic mix moves the fall only from 9.40 to 9.32~points, so the topic mix does not explain it. Who asks separates outcomes far more sharply than what is asked: acceptance runs from 19.8\% for askers below 10~reputation to 66.8\% at 10k and above, a 47-point spread with a gradient in every year, against 17 across topics, and a shift toward newer askers accounts for up to 59\% of the acceptance fall. The largest of seven properties an asker controls accompanies 5~points. The record places the support gap in the toolchain and among newcomers.
\end{abstract}


\section{Introduction}
\label{sec:intro}

Smart contracts on Ethereum~\cite{wood2014ethereum,buterin2014next} run on a public, unchangeable platform where attackers look for bugs. A single misuse of \code{delegatecall}, an unchecked \code{tx.origin} comparison, or a faulty re-entrancy guard~\cite{atzei2017survey} has drained protocols of hundreds of millions of dollars~\cite{chen2020defining,zou2019smart,zhou2023sok}. The developers who write these contracts are diverse and often self-taught, and their number grows faster than official documentation or formal training can follow~\cite{sharma2023mixed,zou2019smart}. Knowing where those developers get stuck is the input that documentation, tooling, and platform design need.

The Ethereum community learns where developers struggle by asking them. Zou et~al.~\cite{zou2019smart} survey and interview contract developers about their challenges, Sharma et~al.~\cite{sharma2023mixed} study their security practices through interviews and a survey, and Abdelaziz et~al.~\cite{abdelaziz2026smart} benchmark six security analyzers on 653 real-world contracts and survey 150 professional developers and auditors on how they use and trust them. Such studies go deep on the theme they choose, but they record stated needs: what a chosen population says when asked about a chosen theme. A survey of practitioners pre-screened for security work can show that false positives and unclear explanations erode trust in analyzers; it cannot show how much of a developer's week goes to problems nobody would call security problems, or whether those problems get solved.

Revealed needs leave a different record. On Ethereum Stack Exchange (\ese), the domain-specific question-and-answer venue, developers post the problem that stopped them, in their own words and on their own initiative, and other developers either resolve it or do not. The archive is a decade of unprompted demand for help paired with the community's supply of it.

Prior work has read that archive only in parts: taxonomies of smart-contract defects and maintenance concerns~\cite{chen2020defining,chen2021maintenance}, a card sort of 2,224~posts on upgradeable contracts~\cite{zhong2025towards}, and a vulnerability scan of the code snippets in answers~\cite{zhou2025smart}. Each examines one sub-topic or one facet of the platform, and none tracks resolution over the platform's full history, a decade that spans the \defi and NFT booms, The Merge, and the rollup era~\cite{werner2022sok,wang2021non,Buterin2022Merge,sguanci2021layer}. Whether the venue still works is open: its annual question volume fell from 8,316 in 2022 to 873 in 2025, and no study has measured what happened to the questions that remained. This gap raises the central question of this paper: \emph{what do Ethereum developers ask for help with, does the venue where they ask still resolve it, and, if not, does the cause lie in the questions developers bring or in how the platform resolves them?}

Answering it requires two things. The first is a topic label for every question, because a fall in resolution means one thing if the tasks developers bring have changed and another if the same tasks stopped being resolved. The second is an analysis that separates those two explanations arithmetically on the complete record rather than a sample of it.

We mine the full \ese archive from \datespan, \nquestions deduplicated questions, and classify every question with \topicgpt~\cite{pham2024topicgpt}, a prompt-based topic-modeling framework, seeded with eight Ethereum-specific topics and free to extend them; the result is \nuniquetopics canonical topics over \ntopicpairs question-topic pairs. A committee of three language models from three providers screens the labels on a stratified sample of 660~questions, and we adjudicate every disputed case by hand, which puts label accuracy between 87.9 and 93.5\%. On the labeled corpus we ask what developers ask for help with and where in the resolution process the platform changed, whether that change reflects a shift in the tasks developers brought or in how the same tasks were resolved, how outcomes vary with who asks and who answers, and what an individual asker controls.

Four answers follow, and they form one story. Developers ask, above all, how to write, deploy, and debug contracts: \emph{Solidity Language} and \emph{Deployment and Tooling} alone hold just over a third of the question-topic pairs, the toolchain is growing, and the security, consensus, and scaling themes that dominate public discussion are thin. The community still answers most of what arrives, but far fewer of those answers are accepted: under each of five estimators of the decade-long change, the answered share fell by 9.5 to 18.6~percentage points and acceptance among the answered by 16.2 to 24.5. Had the later questions kept the earlier topic mix, acceptance would have dropped 9.32~points rather than 9.40, so the change lies within tasks, not in which tasks developers brought. Who is asking matters far more than what is asked: acceptance spans 47~points across five reputation buckets, a gradient present in every year, against 17 across the \nuniquetopics topics; the shift toward newer askers after 2022 accounts for up to 59\% of the fall in acceptance; and the largest association among the seven properties an asker can choose is 5~points.

By locating where developers need help and where the platform fails to give it, our study turns the impression that \ese is dying into a finding that documentation, tooling, and platform designers can act on. This paper makes the following contributions:
\begin{itemize}
  \item A ten-year, corpus-scale record of revealed developer needs on \ese: \nquestions questions in \nuniquetopics validated topics, from a pipeline that labeled the corpus for USD~16.24 and re-runs as the archive grows.
  \item A location for the platform's decline at the closing step of resolution, with an analysis design (a two-stage funnel and direct standardization) that rules out the changing task mix as the cause and that any Q\&A study can reuse.
  \item Evidence that who asks matters more than what is asked: the asker's standing dominates every property of the question, and a shift toward newer askers accompanies up to 59\% of the acceptance fall, which sets the limits of question-writing advice and places the support gap among newcomers.
\end{itemize}

\paragraph*{Data Availability} The labeled corpus, the \topicgpt and judge prompts, the supplementary tables, the analysis and verification scripts, and instructions that regenerate every table and every number in this paper are available in our replication package,\footnote{\url{https://doi.org/10.5281/zenodo.23551228}} so that the platform can be re-baselined as the ecosystem evolves.

\section{Mining and Labeling Ethereum Stack Exchange}
\label{sec:methodology}

To measure what developers ask for help with and whether they get it, per topic and over a decade, we need every question the platform received, a topic label for each one, and evidence that the labels are right.

\subsection{Platform, Outcomes, and Corpus}
\label{sec:method_collection}

\subsubsection{Question Outcomes}
\label{sec:bg_outcomes}
\ese follows the standard Stack Exchange model: users post questions and answers, vote on them, and the asker may mark at most one answer as \emph{accepted}; users accumulate \emph{reputation} from the votes their posts receive. Two properties of these signals shape every interpretation in this paper: acceptance is a community-approval signal, not a correctness or security signal, as Chen et~al.~\cite{chen2019reliable} show for Stack Overflow, and reputation measures past platform activity, not topic expertise. Every question ends in one of three mutually exclusive states: \emph{accepted} if the asker marked an answer as accepted, \emph{orphaned} if it has at least one answer but none accepted, and \emph{never answered} if it has no answers. The \emph{answered} share combines the first two, and \emph{closure} is the accepted share of answered questions. Latency is the time to first answer (TTFA), defined only for answered questions.

\subsubsection{Corpus}
\label{sec:method_cleaning}
Our corpus is every question posted to \ese from \datespan, retrieved from the Stack Exchange Data Explorer~\cite{SEDE2024} and aggregated to one row per question, \nquestions in total. For each question we keep the text, the tags, the creation timestamp, the earliest answer and earliest accepted-answer timestamps, and the reputation of the asker and of the accepted answerer. Two properties of the export bound what the corpus can support: it truncates bodies at 15,000~characters, which affects 0.4\% of questions, and it records reputation as a snapshot at export time, not at posting time. For the reputation analysis we split askers into five buckets, below 10, 10 to 100, 100 to $1\text{k}$, $1\text{k}$ to $10\text{k}$, and $10\text{k}$ and above, each closed on its lower bound. Queries and exclusions are in the replication package.

\subsection{Topic Labeling}
\label{sec:method_topicgpt}

Classical topic models return ranked word lists rather than the categories developers use~\cite{blei2003latent,grootendorst2022bertopic}. We therefore adopt \topicgpt~\cite{pham2024topicgpt}, a prompt-based framework that yields named topics with natural-language definitions through generation, refinement, and assignment stages, and adapt it to this domain. We seed the generation stage with eight Ethereum-specific top-level topics derived from prior taxonomies~\cite{chen2020defining,zou2019smart} (\emph{Security}, \emph{Solidity Language}, \emph{Deployment and Tooling}, \emph{Token Standards}, \emph{Gas Optimization}, \emph{Debugging and Testing}, \emph{Contract Architecture and Upgradeability}, and \emph{Off-chain Integration}). We replace every in-context example with a real \ese post chosen to mark the boundary between two neighboring categories, and constrain the model to propose only generalizable labels while leaving the taxonomy open.

The data reshaped the seeds: seven survive, \emph{Off-chain Integration} is superseded by the finer \emph{Networking}, \emph{Transactions}, and \emph{Storage}, and the corpus contributes categories such as \emph{Consensus Mechanisms}, \emph{Cryptography}, and \emph{Mining}, for \nuniquetopics canonical topics in all. Because a question may concern several topics, we retain multi-label assignments: 12,148~questions carry more than one label, expanding to \ntopicpairs question-topic pairs, and topic-level analyses run on this exploded frame while question-level analyses do not. We run every stage with \code{gpt-4o-mini} in a single pass, at temperature 0 for generation, refinement, and assignment and 0.6 for the correction pass; labeling the corpus cost USD~16.24. The prompts and the post-processing that recovers labels from model output are in the replication package.

\subsection{Label Validation}
\label{sec:method_validation}

The analysis rests on the \topicgpt labels, so we measure how good they are. Manual annotation of \nquestions questions does not scale, and Ahmed et~al.~\cite{ahmed2025can} show that the practical design is hybrid: language models annotate at scale, agreement among several models identifies where automated judgment is trustworthy, and human effort goes to the rest. We follow that design; the committee screens, and manual inspection decides.

\subsubsection{Committee Screening}
We audit a stratified random sample of 30~questions per topic (660 in total). Three language-model judges, \code{gpt-4o} (OpenAI), \code{deepseek-v4-pro} (DeepSeek), and \code{gemini-2.5-pro} (Google), answer one yes-or-no question per assignment: is the assigned topic present in the post? We do not ask for a single best topic, because a developer question often spans several and forcing one choice would count a defensible second label as an error. Each judge receives the same rubric with all \nuniquetopics topic definitions and one present-or-absent rule~\cite{hashemi2024llm}, must cite concrete evidence from the post, at a confidence of at least 0.95, before flagging a label, and runs independently.

Drawing the judges from three organizations limits self-preference and wording sensitivity~\cite{zheng2023judging,he2026llm}, since a flag needs two votes and two judges sit outside the classifier's model family~\cite{verga2024replacing,ahmed2025can}. The committee validates 570 of the 660~assignments and flags 90 (13.6\%), with no abstentions; chance-corrected agreement is Fleiss' $\kappa{=}0.54$, moderate because 85\% of verdicts are ``present'' rather than because the judges disagree, and raw pairwise agreement is 86.7 to 89.5\%. The prompt, per-judge behavior, and per-topic rates are in the replication package.

\subsubsection{Manual Adjudication}
A committee flag does not by itself establish a \topicgpt error, because the judges inherit the reliability limits that make human oversight necessary~\cite{he2026llm,ahmed2025can}, so we manually inspected all 90 flagged assignments. The inspection confirms 43 genuine misclassifications, overturns 10 committee false alarms in which the assigned topic is present, and classifies the remaining 37 as questions that span several topics, so that more than one assignment is defensible. Treating the ambiguous cases as acceptable, the verified label accuracy is $617/660$ (93.5\%); treating them all as errors gives a lower bound of $580/660$ (87.9\%). Both figures assume the 570 committee-validated assignments are correct. The ambiguity is a property of the corpus, not of the classifier: the broad \emph{Contract Architecture and Upgradeability} category, which overlaps with \emph{Solidity Language} and \emph{Transactions}, passes at 50\%, while narrow categories such as \emph{Cryptography}, \emph{Debugging and Testing}, and \emph{Solidity Language} pass at 100\%, which is also why the analysis keeps the multi-label assignments.

\subsection{Analysis Methods}
\label{sec:method_stats}

\begin{table*}[t]
\centering
\footnotesize
\caption{The resolution funnel by year. Answered and never-answered divide by all questions posted that year; $^{*}$accepted and orphaned divide by that year's answered questions. Median TTFA is over answered questions.}
\label{tab:funnel_by_year}
\resizebox{\linewidth}{!}{%
\begin{tabular}{@{}rrrrrrrrr@{}}
\toprule
\textbf{Year} & \textbf{Questions} & \textbf{Answered} & \textbf{Answered\,\%} & \textbf{Never answered} & \textbf{Never answered\,\%} & \textbf{Accepted\,\%$^{*}$} & \textbf{Orphaned\,\%$^{*}$} & \textbf{Median TTFA (h)} \\
\midrule
  2016 & 4,014 & 3,514 & 87.54 & 500 & 12.46 & 59.68 & 40.32 & 2.32 \\
  2017 & 7,816 & 6,768 & 86.59 & 1,048 & 13.41 & 50.84 & 49.16 & 2.69 \\
  2018 & 10,392 & 8,933 & 85.96 & 1,459 & 14.04 & 48.42 & 51.58 & 1.36 \\
  2019 & 4,597 & 3,770 & 82.01 & 827 & 17.99 & 48.78 & 51.22 & 2.15 \\
  2020 & 3,742 & 2,945 & 78.70 & 797 & 21.30 & 50.39 & 49.61 & 2.64 \\
  2021 & 7,488 & 5,605 & 74.85 & 1,883 & 25.15 & 40.77 & 59.23 & 4.50 \\
  2022 & 8,316 & 6,811 & 81.90 & 1,505 & 18.10 & 40.24 & 59.76 & 3.35 \\
  2023 & 4,988 & 4,034 & 80.87 & 954 & 19.13 & 37.75 & 62.25 & 5.08 \\
  2024 & 2,411 & 1,924 & 79.80 & 487 & 20.20 & 37.47 & 62.53 & 8.43 \\
  2025 & 873 & 602 & 68.96 & 271 & 31.04 & 35.22 & 64.78 & 12.83 \\
\bottomrule
\end{tabular}}
\end{table*}

The corpus is every question the platform received in the window rather than a sample of them, so a rate computed over a year, a topic, or a reputation bucket is a property of the whole record, not a sample estimate. We therefore report counts, rates, and differences between them directly, without confidence intervals or significance tests. We report every rate's denominator, in the table that carries it or in the supplementary tables, so that a reader who treats the record as one realization of a process can derive intervals.

To locate a change in the resolution process, we split every question along the two-step path it follows (answered or not; accepted or not, among the answered) and report each step against its own denominator (\Cref{tab:funnel_by_year}). Endpoint contrasts lean on 2025, the smallest year in the record, so we report the decade-long change under five estimators, the 2016-versus-2025 endpoints, the first two years against the last two, the first three against the last three, an ordinary-least-squares (OLS) slope, and a Theil--Sen slope~\cite{sen1968estimates}, each slope multiplied by the nine-year span. The Theil--Sen slope is the median of the 45 pairwise slopes between years, so no single year, neither the 2018 peak nor the 2025 tail, can move it far. We treat the ordering of results as the finding and the magnitude as a range.

To separate a change in the mix of topics from a change in how each topic fares, we use direct standardization~\cite{kitagawa1955components}. With $w_k$ the share of pairs in topic $k$ and $a_k$ its rate, subscripts 0 and 1 for the 2016--2022 and 2023--2025 eras, the observed rates are $\sum_k w_{k0}a_{k0}$ and $\sum_k w_{k1}a_{k1}$ and the later rate rebuilt at the earlier mix is $\sum_k w_{k0}a_{k1}$. The gap between the rebuilt and the observed later rate is what the mix moving did, the gap between the rebuilt and the earlier rate is what happened inside topics, and the two sum to the observed change exactly. Applied to each stage of the funnel separately, the same arithmetic says which stage moved. The split is exact arithmetic on the labeled counts, so it is only as good as the labels.

The reputation analysis compares exact rates across the five asker buckets, measures their spread against the spread across topics, repeats the comparison inside each single year so that unequal time to accumulate reputation cannot produce it, and applies the direct standardization above to the reputation mix of the two eras.

To compare questions with and without a property while controlling for who asked, we compute the difference in acceptance between the two groups inside every stratum of reputation bucket, primary topic (the first label \topicgpt lists), and era. We then take the Mantel--Haenszel weighted mean of those within-stratum differences~\cite{mantel1959statistical}, with weight $n_1 n_0/(n_1+n_0)$ for group sizes $n_1$ and $n_0$. Stratifying on the asker's bucket keeps the comparison from crediting a property with what the asker's standing already explains; strata in which every question shares the property carry no comparison and drop out.

\section{Results}
\label{sec:results}

\subsection{Setup}
\label{sec:setup}

We compute corpus-level rates over the \nquestions questions and per-topic rates over the \ntopicpairs question-topic pairs, report differences in percentage points, and split the decade for before-and-after comparisons at the end of 2022, when code-capable chat assistants became widely available. The replication package rebuilds the labeled corpus from the raw export, and a verification script recomputes every published number from that corpus with code independent of the analysis.

\subsection{Research Questions}
\label{sec:rqs}

The central question that motivates our work is: \emph{what do Ethereum developers ask for help with, does the venue where they ask still resolve it, and, if not, does the cause lie in the questions developers bring or in how the platform resolves them?} We break this question into four sub-questions:
\begin{itemize}
  \item \rqueue{1}: What do Ethereum developers ask for help with, and at which of the two stages of resolution, a question receiving an answer or an answer being accepted, did \ese change between 2016 and 2025?
  \item \rqueue{2}: Is that change attributable to a shift in the mix of development tasks developers asked about, or to a change in how the same tasks were resolved?
  \item \rqueue{3}: How do resolution outcomes vary with asker reputation, and how have the reputation profiles of askers and accepted answerers changed over the same period?
  \item \rqueue{4}: Which question properties under the asker's control are associated with acceptance, and how do those associations compare in size with the asker-reputation gradient?
\end{itemize}

\subsection{Revealed Demand and the Stage of Decline (\rqueue{1})}
\label{sec:rq1}

\begin{figure}[!htb]
\centering
\includegraphics[width=\linewidth]{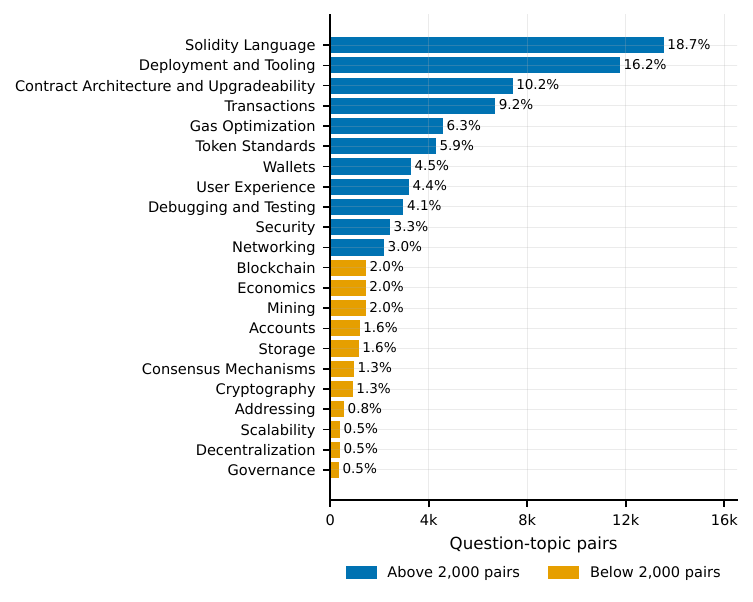}
\caption{The \nuniquetopics canonical topics by question-topic pair count and share of all pairs. Blue marks the eleven topics above 2,000~pairs.}
\label{fig:topic_inventory}
\end{figure}

\subsubsection{What Developers Ask For}
Demand on \ese is concentrated and practical. \emph{Solidity Language} and \emph{Deployment and Tooling} alone account for 34.9\% of question-topic pairs, the top five topics for 60.6\%, and the eleven topics above 2,000~pairs for 85.9\% (\Cref{fig:topic_inventory}). The subjects that dominate public discussion of Ethereum are thin in the developer archive: \emph{Security} is 3.3\% of pairs, \emph{Consensus Mechanisms} 1.3\%, and \emph{Scalability} and \emph{Governance} 0.5\% each. The mix shifts across the decade: \emph{Contract Architecture and Upgradeability} roughly doubles its share of pairs, from 8.0\% in 2016 to 15.0\% in 2025, \emph{Deployment and Tooling} grows from 14.2\% to 18.1\% and \emph{User Experience} from 2.8\% to 6.9\%, while \emph{Networking} falls from 6.2\% to 1.8\% and \emph{Mining} from 6.0\% to 0.2\%. The doubling rests on the one category the audit validates at only 50\% (\Cref{sec:method_validation}), so its share is the least certain in the inventory; its closest neighbor, \emph{Solidity Language}, is flat (11.7\% to 10.7\%), so the growth is not a boundary shift between the two.

\begin{table*}[t]
\centering
\footnotesize
\caption{Demand and supply for the eleven topics above 2,000 question-topic pairs: share of pairs and its shift from 2016--2022 to 2023--2025, answered and acceptance rates, and median time to first answer over answered pairs.}
\label{tab:rq1_demand_supply}
\begin{tabular}{@{}lrrrrr@{}}
\toprule
\textbf{Topic} & \textbf{Pairs\,\%} & \textbf{Shift (pp)} & \textbf{Answered\,\%} & \textbf{Accepted\,\%} & \textbf{TTFA (h)} \\
\midrule
  Solidity Language & 18.7 & -2.0 & 85.6 & 44.3 & 2.0 \\
  Deployment and Tooling & 16.2 & 1.5 & 79.7 & 33.8 & 5.7 \\
  Contract Architecture and Upgradeability & 10.2 & 2.0 & 83.8 & 38.5 & 3.0 \\
  Transactions & 9.2 & 1.4 & 79.1 & 33.8 & 3.4 \\
  Gas Optimization & 6.3 & -0.2 & 79.8 & 37.2 & 2.7 \\
  Token Standards & 5.9 & -0.7 & 84.7 & 37.3 & 2.4 \\
  Wallets & 4.5 & 0.0 & 78.3 & 26.9 & 3.4 \\
  User Experience & 4.4 & 2.1 & 77.9 & 28.7 & 4.7 \\
  Debugging and Testing & 4.1 & 1.9 & 76.9 & 36.4 & 5.5 \\
  Security & 3.3 & 0.3 & 83.1 & 39.6 & 2.4 \\
  Networking & 3.0 & -1.8 & 73.3 & 32.5 & 6.4 \\
\bottomrule
\end{tabular}
\end{table*}

\subsubsection{Demand Against Supply}
\Cref{tab:rq1_demand_supply} sets each main topic's share of demand beside how the community supplies it. \emph{Solidity Language} is the largest topic and the best served: 85.6\% of its pairs are answered, 44.3\% accepted, at a median of 2.0~hours. \emph{Deployment and Tooling} is the second largest, still growing (+1.5~points between eras), and served slowly, at 5.7~hours and 33.8\% accepted. Three of the four shares that grew most between eras, \emph{Deployment and Tooling} (+1.5), \emph{User Experience} (+2.1), and \emph{Debugging and Testing} (+1.9), are the three slowest topics after \emph{Networking}, which is the slowest of the eleven (6.4~hours) and the least answered (73.3\%); only \emph{Contract Architecture and Upgradeability} (+2.0) grows and resolves at the corpus pace. \emph{Security}, at 3.3\% of pairs, is answered faster and accepted more often than the corpus as a whole. Demand is moving toward topics the community serves slowly, the gap the rest of the paper measures: whether it opened because the questions changed (\rqueue{2}), whom it falls on (\rqueue{3}), and what an asker can do about it (\rqueue{4}).

\subsubsection{Where Resolution Failed}
Of \nquestions questions, 44,906 (82.2\%) drew an answer and 9,731 never did; of the answered, 20,668 (46.0\%) reached an accepted answer and 24,238 (54.0\%) were orphaned. Separating the stages over time shows that they moved by different amounts (\Cref{tab:funnel_by_year}). The answered share fell from 87.5\% in 2016 to 69.0\% in 2025, a drop of 18.6~percentage points. Among answered questions, the accepted share fell from 59.7\% to 35.2\%, a drop of 24.5, so that by 2025 nearly two in three answered questions were left orphaned (64.8\%, against 40.3\% in 2016). Both endpoints depend on 2025, the smallest year at 873~questions, so \Cref{tab:rq1_trajectory} reports the change under four further estimators of the decade-long trend. The answered share falls by 9.5 to 18.6~points and closure by 16.2 to 24.5 depending on the estimator; all five agree on direction, and every one puts the closure fall above the answering fall.

\begin{table*}[t]
\centering
\footnotesize
\caption{The decade-long change in each funnel stage under five estimators (percentage points). Slopes are multiplied by the nine-year span so that every column is a change over the decade.}
\label{tab:rq1_trajectory}
\begin{tabular}{@{}lrrrrr@{}}
\toprule
\textbf{Measure} & \textbf{2016 vs 2025} & \textbf{First 2 vs last 2} & \textbf{First 3 vs last 3} & \textbf{OLS slope} & \textbf{Theil--Sen slope} \\
\midrule
  Answered (of all questions) & -18.59 & -12.69 & -10.15 & -13.33 & -9.46 \\
  Accepted (of answered) & -24.46 & -18.91 & -16.16 & -21.94 & -20.35 \\
\bottomrule
\end{tabular}
\end{table*}

\subsubsection{Latency and the End of the Record}
The wait for a first answer lengthened from a median of 2.32~hours in 2016 to 12.83 in 2025, measured over answered questions only (\Cref{tab:funnel_by_year}); the median is the right summary because the mean is 184 times larger. The wait was shortest in 2018, the platform's largest year (10,392~questions, a median of 1.36~hours), so measured from the peak the 2025 wait is 9.4 times longer. Truncation at the end of the record cannot explain the 2025 figures: 97.2\% of all acceptances arrive within the shortest observation window any question received, 104~days, which caps the room for the 2025 acceptance rate to rise at 0.70~points.

\begin{roundbox}
\textbf{\rqueue{1}:} Developers ask for help with writing, deploying, and debugging contracts, not with the consensus, scaling, and governance themes that dominate public discussion, and demand is moving toward toolchain and debugging topics that the community serves slowly. The platform weakened at both stages of resolution but unevenly: answering fell by 9.5 to 18.6~points and closure among answered questions by 16.2 to 24.5 across five estimators of the decade. Every one puts the second fall above the first.
\end{roundbox}

\subsection{Task Mix Versus Resolution of the Same Tasks (\rqueue{2})}
\label{sec:rq2}

\Cref{sec:rq1} shows demand moving toward slowly served topics. If the growing topics are also accepted less often, acceptance would fall without anyone answering worse, and every later claim in this paper would describe the questions, not how they were resolved.

\subsubsection{The Spread the Mix Could Exploit}
Topics do differ enough for the mix to matter. Over all \nuniquetopics topics, measured against all of a topic's pairs, acceptance ranges from 26.9\% for \emph{Wallets} to 44.3\% for \emph{Solidity Language}, a spread of 17.3~points, the answered share from 73.3\% for \emph{Networking} to 85.6\% for \emph{Solidity Language}, and median latency from 1.6~hours for \emph{Addressing} to 7.4 for \emph{Consensus Mechanisms} (\Cref{tab:rq1_demand_supply} carries the eleven largest; the profile of all \nuniquetopics topics is in the replication package). The gaps hold at each stage taken alone: \emph{Networking}, \emph{Debugging and Testing}, \emph{Consensus Mechanisms}, and \emph{User Experience} are never answered most often (22.1 to 26.7\%, against 14.4\% for \emph{Solidity Language}).

\begin{table*}[t]
\centering
\footnotesize
\caption{Each rate in the two eras, the later rate rebuilt at the 2016--2022 topic mix, and the split of the change into the part the mix moving explains and the part that happened inside topics.}
\label{tab:rq2_standardization}
\resizebox{\linewidth}{!}{%
\begin{tabular}{@{}lrrrrrr@{}}
\toprule
\textbf{Outcome} & \textbf{2016--2022\,\%} & \textbf{2023--2025\,\%} & \textbf{Later at earlier mix\,\%} & \textbf{Change (pp)} & \textbf{Mix (pp)} & \textbf{Within topics (pp)} \\
\midrule
  Accepted (of all pairs) & 38.46 & 29.06 & 29.14 & -9.40 & -0.08 & -9.32 \\
  Answered (of all pairs) & 81.91 & 78.29 & 77.86 & -3.61 & 0.43 & -4.05 \\
  Accepted (of answered pairs) & 46.96 & 37.12 & 37.32 & -9.84 & -0.20 & -9.64 \\
\bottomrule
\end{tabular}}
\end{table*}

\subsubsection{Acceptance at the Earlier Mix}
The mix nevertheless explains almost none of the decline. Acceptance over question-topic pairs fell from 38.5\% in 2016--2022 to 29.1\% in 2023--2025, a drop of 9.40~points; the later rate rebuilt at the earlier era's topic mix is 29.14\%, a drop of 9.32 (\Cref{tab:rq2_standardization}). The mix moving accounts for 0.08 of the 9.40~points, under 1\% of the fall, because the topics that grew are not systematically better or worse resolved than the topics that shrank.

\subsubsection{The Stage That Moved}
The same standardization applied to each stage of the funnel separately locates the residual (\Cref{tab:rq2_standardization}). The mix term is small at both stages ($+0.4$ and $-0.2$~points), and the within-topic movement is $-9.6$~points on the acceptance step against $-4.0$ on the answering step. The stage where the asker closes the question out moved more than twice as far as the stage where the community supplies an answer, and it is the same asymmetry \Cref{sec:rq1} finds across the corpus. Because the labels follow the task a developer is performing, the result rules out a shift in the mix of tasks as these labels measure it; a shift in subject matter within a task, which the labels cannot see, remains possible, and \Cref{sec:rq3} measures the shift in who asks.

\begin{roundbox}
\textbf{\rqueue{2}:} The mix of tasks developers brought explains almost none of the decline: at the earlier mix, acceptance would have fallen 9.32~points instead of the observed 9.40. The change is within tasks, and within tasks it is concentrated on the closing step.
\end{roundbox}

\subsection{Asker Reputation and the Answerer Pool (\rqueue{3})}
\label{sec:rq3}

\begin{table*}[t]
\centering
\footnotesize
\caption{Outcomes by asker reputation, over all questions with a recorded reputation. The last column conditions on an answer existing.}
\label{tab:rq3_askers}
\begin{tabular}{@{}lrrrrr@{}}
\toprule
\textbf{Asker reputation} & \textbf{Questions} & \textbf{Accepted\,\%} & \textbf{Never ans.\,\%} & \textbf{Within 24h\,\%} & \textbf{Within 24h of answered\,\%} \\
\midrule
  $<$10 & 7,158 & 19.80 & 19.01 & 62.77 & 77.51 \\
  10--100 & 17,129 & 27.89 & 22.28 & 58.30 & 75.02 \\
  100--1k & 22,722 & 45.12 & 15.64 & 65.54 & 77.69 \\
  1k--10k & 6,034 & 53.60 & 13.84 & 67.09 & 77.86 \\
  $\geq$10k & 1,198 & 66.78 & 10.60 & 69.37 & 77.59 \\
\bottomrule
\end{tabular}
\end{table*}

The decline sits inside tasks; the next question is who asks and who is left to answer.

\subsubsection{Who Is Served}
Asker reputation separates outcomes more sharply than anything else we measure. Acceptance climbs at every step of the five-bucket scale, from 19.8\% for askers below 10~reputation to 66.8\% for those at $10\text{k}$ and above (\Cref{tab:rq3_askers}): a spread of 47.0~points and a ratio of $3.4\times$. That spread is 2.7 times the 17.3-point spread across the 22~topics; the top bucket holds 1,198~questions, 2.2\% of the frame.

The advantage lies in getting an answer at all rather than in getting one sooner: about one in five questions from askers below 100~reputation never receives an answer, against one in ten from askers at $10\text{k}$ and above. Within 24~hours, 62.8\% of questions from the lowest bucket are answered against 69.4\% from the highest. Among questions that did receive an answer, the within-24-hour share runs from 75.0\% to 77.9\%, with the top bucket within 0.1~points of the bottom.

\begin{table*}[t]
\centering
\footnotesize
\caption{Acceptance by asker reputation bucket inside each year, as a percentage of that year's questions in the bucket. Bucket sizes are in the replication package.}
\label{tab:rq3_gradient_by_year}
\begin{tabular}{@{}rrrrrrrr@{}}
\toprule
\textbf{Year} & \textbf{$<$10} & \textbf{10--100} & \textbf{100--1k} & \textbf{1k--10k} & \textbf{$\geq$10k} & \textbf{Top $-$ bottom (pp)} & \textbf{Rising steps (of 4)} \\
\midrule
  2016 & 40.0 & 22.3 & 48.0 & 64.3 & 77.7 & 37.7 & 3 \\
  2017 & 25.7 & 25.6 & 50.3 & 57.3 & 66.7 & 41.0 & 3 \\
  2018 & 29.3 & 32.2 & 47.6 & 46.0 & 46.8 & 17.5 & 3 \\
  2019 & 24.8 & 28.6 & 47.8 & 50.7 & 57.7 & 32.9 & 4 \\
  2020 & 23.0 & 32.8 & 45.3 & 46.7 & 69.3 & 46.4 & 4 \\
  2021 & 14.0 & 23.7 & 38.2 & 47.2 & 71.8 & 57.9 & 4 \\
  2022 & 21.2 & 28.6 & 41.4 & 47.6 & 69.9 & 48.7 & 4 \\
  2023 & 17.2 & 25.6 & 41.1 & 51.2 & 61.6 & 44.5 & 4 \\
  2024 & 17.3 & 31.6 & 40.5 & 43.7 & 62.1 & 44.8 & 4 \\
  2025 & 9.3 & 21.4 & 39.4 & 40.9 & 66.1 & 56.8 & 4 \\
\bottomrule
\end{tabular}
\end{table*}

\subsubsection{The Gradient Inside Each Year}
The export-time reputation snapshot gives a 2016 asker a decade to accumulate reputation and a 2025 asker months, so the pooled gradient may reflect accumulation time. Compared only with askers who posted in the same year, it survives (\Cref{tab:rq3_gradient_by_year}): the top-minus-bottom spread runs from 17.5 to 57.9~points across the ten years, exceeds 30 in nine of them, and rises at all four steps in seven. If accumulation after the question drove it, the spread would be widest for 2016 and narrowest for 2025; it is 37.7 in 2016 and 56.8 in 2025. The 396~questions whose asker has no recorded reputation fare better than the corpus, 47.0\% accepted against 37.8\%, so excluding them cannot inflate the gradient.

\subsubsection{The Asker Mix}
The gradient bears on the decline because the askers moved down the scale: those below 100~reputation posted 41.0\% of questions before 2023 and 65.8\% after. At the earlier reputation mix, standardized as \Cref{sec:rq2} did for topics, later acceptance would be 35.3\% instead of the observed 29.7\%, so the mix accounts for 5.6 of the 9.5-point fall per question (59\%), and acceptance fell inside all five buckets, by 3.9~points on average. Among answered questions it covers 6.4 of the 10.0-point closure fall and 1.2 of the 3.3-point answering fall, so inside buckets closure still fell further than answering (3.6 against 2.1~points). The export-time snapshot places recent askers in lower buckets than their eventual standing, so these mix terms overstate what a real shift in who asks explains (bucket shares and standardization in the replication package).

\begin{figure}[!htb]
\centering
\includegraphics[width=\linewidth]{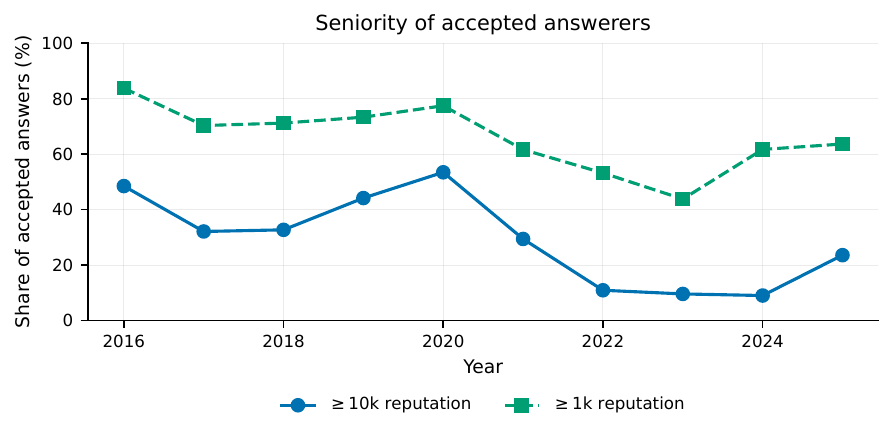}
\caption{Share of accepted answers written by users at or above $10\text{k}$ and $1\text{k}$ reputation, by year.}
\label{fig:rq3_answerers}
\end{figure}

\subsubsection{Who Answers}
The pool writing accepted answers grew less senior, then partly recovered (\Cref{fig:rq3_answerers}). The share of accepted answers from users at or above $10\text{k}$ reputation fell from 48.5\% in 2016 to a trough of 9.0\% in 2024 before returning to 23.6\% in 2025, and the median accepted-answerer reputation from 9,693 to 1,641. The broader pool at or above $1\text{k}$ follows the same path, from 83.8\% of accepted answers in 2016 to 43.8\% in 2023 and 63.7\% in 2025. The late years rest on small bases, 212 accepted answers in 2025 against 2,039 in 2016, so the partial recovery is less certain than the decade-long fall. Accumulation cannot alone account for the pattern, because the median accepted-answerer reputation peaks at 12,925 in 2020, above its 2016 value.

\subsubsection{The Expertise Gap on Matched Populations}
Paired within each accepted question, the answerer holds a median $13.3\times$ the asker's reputation over 20,337~questions, and is ahead in 75.3\% of them. Dividing the median answerer reputation (3,408) by the median asker reputation over all questions (113) gives $30.2\times$, but it compares different question sets: answerer reputation exists only for accepted questions, whose askers have a higher median (182).

\begin{roundbox}
\textbf{\rqueue{3}:} Who asks separates outcomes two to three times more sharply than what is asked about, the gradient appears in every single year, and the advantage shows in whether a question is answered at all, with almost no difference in how soon. After 2022 the askers moved down the scale, a shift behind up to 59\% of the fall in acceptance, and acceptance also fell inside every bucket. The pool writing accepted answers lost much of its seniority, recovering part of it in 2024--2025 on a small base, and within a question the answerer holds about $13\times$ the asker's reputation.
\end{roundbox}

\subsection{Asker-Controlled Question Properties (\rqueue{4})}
\label{sec:rq4}

\begin{table*}[t]
\centering
\footnotesize
\caption{Association between asker-controlled properties and acceptance (percentage points). Adjusted differences are Mantel--Haenszel means within reputation bucket, primary topic, and era; the last column counts reputation buckets with a positive difference.}
\label{tab:rq4_effects}
\begin{tabular}{@{}lrrrrl@{}}
\toprule
\textbf{Feature} & \textbf{Acc.\ with\,\%} & \textbf{Acc.\ without\,\%} & \textbf{Raw diff (pp)} & \textbf{Adjusted diff (pp)} & \textbf{Buckets positive} \\
\midrule
  Includes a code snippet & 41.50 & 33.10 & 8.40 & 5.00 & 4/5 \\
  Body longer than median & 37.93 & 37.59 & 0.33 & 0.08 & 3/5 \\
  Four or more tags & 39.45 & 37.03 & 2.42 & 2.08 & 5/5 \\
  Title longer than median & 37.46 & 38.04 & -0.58 & -0.28 & 1/5 \\
  Names a compiler or language version & 41.30 & 37.22 & 4.07 & 2.56 & 4/5 \\
  Names a development framework & 38.15 & 37.65 & 0.50 & 0.98 & 4/5 \\
  Quotes an error message or trace & 39.61 & 37.53 & 2.08 & 1.46 & 4/5 \\
\bottomrule
\end{tabular}
\end{table*}

The forces described so far are ones an individual developer cannot change; what remains is the question itself.

\subsubsection{Raw and Adjusted Associations}
Seven properties an asker decides when writing a post are visible in the export: whether the post contains a code snippet, whether it carries four or more tags, and whether its body and title are longer than the corpus median; and, read from the body text with documented regular expressions, whether it names a compiler or language version, names a development framework, and quotes an error message or trace. The three text markers are present in 13.1\%, 21.9\%, and 10.9\% of questions.

Their raw associations with acceptance are confounded by who asks, so \Cref{tab:rq4_effects} reports both the raw difference and the Mantel--Haenszel difference within reputation bucket, primary topic, and era (of the 217 non-empty strata, 129 to 205 contribute per property). Adjustment matters most for the property that looks most useful: a code snippet accompanies 8.4~points higher raw acceptance, of which 5.0 survives, so roughly 40\% of the apparent advantage is explained by who was posting, about what, and when, not by the code. A version number accompanies 2.6~points after adjustment, four or more tags 2.1, an error trace 1.5, and a framework name 1.0; body and title length accompany 0.1 and $-0.3$, which we read as no association.

\subsubsection{Direction Across the Reputation Scale}
Direction matters more than magnitude, because a property that helps one kind of asker and hurts another cannot be given as advice (per-bucket differences are in the replication package). Tag count is the only property that points the same way in all five reputation buckets. Code reverses at the top: askers at $10\text{k}$ and above who include a snippet are accepted less often than those who do not, by 3.1~points, against a 5.8 to 9.1-point advantage in the four buckets below. The version and error-trace markers follow the code snippet, with 1.9 to 7.2~points higher acceptance in the four lower buckets and 12.8 and 10.2~points lower in the top one. The framework marker also turns negative at the top ($-8.5$). Body length reverses in the other direction, accompanying higher acceptance among the lowest-reputation askers and lower among the highest, so neither ``keep it short'' nor ``write more'' survives as general advice. The adjusted differences are associations in an archival record, not effects of changing a question.

\begin{roundbox}
\textbf{\rqueue{4}:} Little of the outcome is under the asker's control. Of seven properties an asker chooses, only tag count points the same way across the whole reputation scale, at about 2~points. The largest association, a code snippet at 5.0~points after adjustment, is one ninth of the 47-point reputation spread and reverses for the most established askers, as do the version, framework, and error-trace markers.
\end{roundbox}

\section{Discussion}
\label{sec:discussion}

\subsection{What the Record Means for the Ethereum Community}
\label{sec:disc_implications}

\emph{Grouped by the kind of work a question concerns, the topics fall into four clusters that call for different effort} (membership and per-topic rates in the replication package). \emph{Toolchain and integration}, seven topics led by \emph{Deployment and Tooling}, \emph{Transactions}, and \emph{Wallets}, holds 40.9\% of pairs, gained 6.0~points after 2022, and is the slowest to answer (4.35~hours). The \emph{language core}, led by \emph{Solidity Language}, is the fastest (2.21~hours) and lost the least acceptance (42.7\% to 34.9\%), \emph{contract design and security} lost the most (41.1\% to 28.4\%), and nine \emph{protocol and ecosystem} topics fell from 13.9\% to 9.1\% of pairs.

\emph{Tool builders should start with toolchain and integration.} Its questions turn on a compiler version, a framework, a network, or a wallet state that an answerer must reconstruct, and \emph{Wallets} lost more acceptance than any other main topic (29.2\% to 15.7\%), so environment-aware diagnostics (\Cref{sec:disc_newcomers}) and wallet and front-end SDK documentation target the largest, slowest need. For researchers, these questions offer a benchmark for assistants on environment-dependent failures. \emph{Debugging and Testing}, often never answered yet barely less accepted after 2022 (2.4~points), suggests that the bottleneck is reaching someone who can reproduce a problem.

\emph{Design and security checks belong inside the workflow.} \emph{Contract Architecture and Upgradeability} lost 12.9~points of acceptance (40.9\% to 28.1\%) while growing, which motivates upgrade-safety tooling, such as proxy-pattern and storage-layout checks. \emph{Security}, the theme of recent surveys~\cite{sharma2023mixed,abdelaziz2026smart}, kept its 3.3\% share but lost its advantage: 41.7\% accepted before 2023, 29.7\% after, against 29.1\% for all pairs. Incident-class mistakes~\cite{atzei2017survey,chen2020defining}, such as \code{tx.origin} authorization or \code{delegatecall} misuse, arrive as language, token, or architecture questions, so analyzers~\cite{abdelaziz2023smart,abdelaziz2026smart} and courses reach most developers only in the compiler, the framework, and the documentation. Runtime monitors~\cite{abdelaziz2026txlens} act only after deployment.

\emph{The language core is the most reusable knowledge; the protocol topics are fading.} The best-served cluster suits curated documentation and retrieval sources for assistants once screened for security (\Cref{sec:disc_newcomers}), while after 2022 \emph{Mining} has 37~pairs and \emph{Networking} 184, too few for studies of node operation or consensus.

\emph{Onboarding capacity is falling.} Newcomers, the developers the ecosystem most needs to keep, now post most of the questions and are the ones the venue serves worst: the reputation gradient holds inside every year, one newcomer question in five is never answered against one in ten from established askers, and the senior answerer pool thinned (\Cref{sec:rq3}). The venue is losing what a newcomer needs, an expert who answers and a question that gets closed, and is becoming an archive of 20,668 accepted answers whose live layer needs deliberate support.

\subsection{What Changed When Assistants Arrived}
\label{sec:disc_decay}

Our two eras meet in late 2022, with the arrival of code-capable chat assistants~\cite{chen2021evaluating}. Volume and waiting broke at that boundary: questions fell from 8,316 in 2022 to 873 in 2025 and median TTFA rose every year, from 3.35 to 12.83~hours. Closure did not: among answered questions it fell 19.4~points from 2016 to 2022 and a further 5.0 by 2025, and the 10k-and-above share of accepted answers had already fallen from 53.5\% in 2020 to 10.9\% in 2022 (\Cref{fig:rq3_answerers}). Assistants may explain why fewer questions arrive and why they wait longer, but not a decline in closure and a thinning of the expert pool that began years earlier. Even the volume fall has a rival: volume had already dropped 64\% in the 2018--2020 crypto-market downturn, and the 2022--2023 downturn coincides with the second fall, so the corpus cannot separate it from the assistant effect measured on Stack Overflow~\cite{del2024large} without an external activity index.

\emph{After 2022 the archive is a filtered sample.} Each later question is a need an assistant did not meet (\Cref{sec:threats_external}), so falling volume need not mean falling need, and a question left unanswered or orphaned is one not resolved on \ese, not necessarily unresolved. If assistants absorb the easy questions, the remainder should tilt toward the tasks that fare worse; across the \nuniquetopics tasks it does not: the post-2022 topic mix, taken alone, raises the answered share by 0.4~points and lowers acceptance by 0.08 (\Cref{tab:rq2_standardization}). The questions and the askers did change (\Cref{sec:rq3}): all seven asker-controlled properties are more common after 2022, with an error trace in 15.1\% of questions against 10.2\% before and a code snippet in 57.8\% against 55.1\% (per-property shares in the replication package). The later questions carry more of the context that accompanies acceptance, yet closure kept falling, for reasons the archive cannot see.

\subsection{What to Build for Newcomers}
\label{sec:disc_newcomers}

\emph{Capture the context at the source.} Nothing an asker controls accompanies more than 5.0~points (\Cref{sec:rq4}), but below 10k~reputation a code snippet, a version, and an error trace accompany 1.9 to 9.1, and a framework's issue command, a compiler, or an IDE can attach them. A template that captures versions, network, and error trace is the intervention this study motivates most directly; only a controlled deployment can test whether it changes the outcome.

\emph{Route for reach, not speed.} The reputation advantage lies in being answered at all, not sooner: among answered questions, the share answered within 24~hours differs by under 3~points across the buckets (\Cref{sec:rq3}). The lever is matching rather than latency, through a first-question queue, mentors, or routing to experts by topic, and its metric is the never-answered rate by reputation bucket.

\emph{Put experts where assistants fail, and screen what the archive feeds them.} Assistants answer in seconds, but 52\% of ChatGPT's answers to Stack Overflow questions contain incorrect information~\cite{kabir2024stack}, so our findings motivate, but do not test, a hybrid in which a model drafts an answer from the archive through retrieval-augmented generation~\cite{lewis2020retrieval} and a high-reputation user verifies it. For Ethereum the retrieval step needs a security check, because snippets on \ese contain vulnerabilities~\cite{zhou2025smart} and vulnerable Q\&A snippets reach deployed contracts~\cite{weiss2024analyzing}, so an assistant that serves accepted answers to a newcomer who cannot judge them passes those flaws on. The closure result also asks the platform for a resolution signal that does not depend on the asker returning, since after 2022 an asker who stops returning may have solved the problem elsewhere, and for incentives that keep senior answerers.

\subsection{Implications for Mining Software Repositories}
\label{sec:disc_msr}

Four choices in this study transfer to any Q\&A archive. Report resolution in stages, each against its own denominator: the single acceptance rate, 52.2\% to 24.3\%, hides that answers still arrive. Standardize every composition that moved before attributing a change: here the topic mix explains under 1\% of the acceptance fall and the asker mix up to 59\%, and snapshot confounds need comparison inside each year and paired populations. Validate language-model labels per category: labeling cost USD~16.24 and the overall accuracy is 87.9 to 93.5\%, yet one broad category passes at 50\%. Date the assistant filter: data after late 2022 is a selected sample of developer needs, so a study should report its share of the corpus (15.1\% here) and read volume with resolution against the earlier trend.

\section{Threats to Validity}
\label{sec:threats}

\subsection{Internal Validity}
\label{sec:threats_internal}

The audit checks that each assigned topic is present, so it measures label precision, not missed topics. Its accuracy of 87.9 to 93.5\% (\Cref{sec:method_validation}) assumes the 570 committee-validated assignments are correct, which we did not re-check, and \emph{Contract Architecture and Upgradeability} passes at only 50\% on 30 sampled assignments, so claims resting on it are less certain than the overall rate suggests. The result that depends most on the labels is the standardization in \Cref{sec:rq2}: a label error moves its mix term only if it is aligned with both era and topic acceptance, and the term it would have to move is 0.08 of 9.40~points. We did not re-run the standardization under perturbed labels or at another era boundary.

Per-topic rates run on the exploded question-topic frame, where the 22.2\% of multi-topic questions contribute one row per topic. No descriptive rate needs independence to be exact, and the adjustment in \Cref{sec:rq4} runs on the disjoint primary-topic frame, whose primary label follows \topicgpt's unfixed label order, which we did not vary. The topic standardization cannot separate askers changing what they accept from questions inside each topic getting harder, and the reputation standardization in \Cref{sec:rq3} inherits the export-time snapshot, which understates recent askers' standing and so inflates its mix term. The export carries no user identifiers, so nothing in this paper separates answers that stopped deserving acceptance from askers who stopped returning or solved the problem elsewhere.

Without sampling intervals, the contrasts most exposed to noise are the smallest, body and title length at 0.1 and $-0.3$~points, which we already read as no association. The three text markers in \Cref{sec:rq4} record that a version, framework, or error string is mentioned, not that the report is complete.

\subsection{External Validity}
\label{sec:threats_external}

\ese attracts English-speaking developers who post in public, so its users do not represent all smart-contract developers, and a question on \ese is a need that outlasted documentation, search, and, after late 2022, an assistant. The archive under-represents problems developers solve alone. The corpus stops at the end of 2025, but late acceptances can raise the 2025 rate by at most 0.70~points (\Cref{sec:rq1}), against a fall from 52.2\% to 24.3\% across the decade. The link between the volume collapse and language models (\Cref{sec:disc_decay}) is a timing correlation, not a causal claim.

\subsection{Construct Validity}
\label{sec:threats_construct}

Acceptance measures community approval, not correctness (\Cref{sec:bg_outcomes}), and because reputation measures past activity rather than expertise, the gradient may mean that long-time users write better questions rather than that the community treats them better. Reputation is also an export-time snapshot: the within-year gradient (\Cref{tab:rq3_gradient_by_year}) shows that unequal accumulation time does not produce it, but it does not rule out that askers whose questions were accepted stayed and earned reputation afterwards, so part of the gradient, and of the asker-mix term, may run from outcome to standing. The \emph{Security} label covers only questions developers frame as security questions, so it may under-report the security signal; measuring that gap would need a topic-blind classifier such as PUMiner~\cite{le2020puminer}. One judge, \code{gpt-4o}, shares \topicgpt's model family and may favor its labels~\cite{zheng2023judging}; a flag needs two of three judges, and this judge approves 84.4\% of assignments against Gemini's 91.5\%, so it is not the most lenient.

\section{Related Work}
\label{sec:related}

\subsection{Stated Needs: Surveys and Interviews}
\label{sec:rw_surveys}

The Ethereum community's picture of developer needs comes from asking developers. Zou et~al.~\cite{zou2019smart} survey and interview smart-contract developers and identify weak tooling, immature language features, and limited online resources as the main pain points, and Sharma et~al.~\cite{sharma2023mixed} interview and survey developers about their security practices. Abdelaziz et~al.~\cite{abdelaziz2026smart} pair a benchmark of six security analyzers on 653 real-world contracts with a survey of 150 professional developers and auditors, and find that excessive false positives (33.3\%) and unclear explanations (18.7\%) are the main reasons practitioners ignore analyzer warnings. These studies reach developers directly and can ask why, which an archive cannot.

We read the complementary record: a decade of questions developers chose to post, labeled by task and paired with the community's response. The two views agree where they overlap, since the toolchain pain points those surveys report fall in the largest and slowest-answered cluster we find. The archive adds what these surveys do not measure: how little of the daily struggle is framed as security, and how much of the outcome depends on who asks.

\subsection{Mining Ethereum Stack Exchange}
\label{sec:rw_ese}

Chen et~al. define a taxonomy of 20 smart-contract defects from \ese posts and on-chain contracts~\cite{chen2020defining} and catalog the maintenance concerns of deployed contracts~\cite{chen2021maintenance}. Zhong et~al.~\cite{zhong2025towards} are the closest in aim to our study: the authors extract 2,224~posts on upgradeable contracts from \ese by keyword and manual filtering and derive 13 development concerns by open card sorting~\cite{spencer2009card}. The post-based studies work at hundreds to a few thousand posts with manual coding, which allows neither long-term analysis nor a view of the platform as a whole. We analyze \nquestions questions over a decade, and that scale surfaces categories the sampled studies do not report, such as \emph{Consensus Mechanisms} and \emph{Networking}.

Zhou et~al.~\cite{zhou2025smart} shift from topics to answer reliability and report that 11.18\% of contract-level and 4.06\% of function-level snippets on \ese contain vulnerabilities, and Weiss et~al.~\cite{weiss2024analyzing} trace vulnerable Stack Overflow Solidity snippets into 17,852 deployed contracts. They study the answer side of the platform; we supply the question-side baseline they presuppose, including where the security signal sits in the taxonomy (\Cref{sec:disc_implications}).

\subsection{Mining Developer Q\&A Platforms}
\label{sec:rw_qa}

Beyond Ethereum, Yang et~al.~\cite{yang2016security}, Lopez et~al.~\cite{lopez2019anatomy}, and Le et~al.~\cite{le2021large,le2020puminer} mine security questions on Stack Overflow at scale, and Chen et~al.~\cite{chen2019reliable} show that the community's scoring does not tell secure code from insecure code, a caveat our acceptance-based outcomes inherit (\Cref{sec:threats}). Early studies of smart-contract discussion on Stack Overflow and Medium report a large unanswered share and limited explicit security discussion~\cite{ayman2020smart,ayman2019smart}, and Alahi et~al.~\cite{alahi2019identifying} observe with latent Dirichlet allocation~\cite{blei2003latent} a shift away from mining toward development; we see all three at full scale (\Cref{sec:rq1,sec:rq2}).

Asaduzzaman et~al.~\cite{asaduzzaman2013answering} attribute most unanswered Stack Overflow questions to questions that fail to reach an expert or are too short, unclear, or too specific; the never-answered stage of \Cref{sec:rq1} measures the size of that outcome on \ese without classifying its causes. Calefato et~al.~\cite{calefato2018ask} derive guidelines for writing Stack Overflow questions from the properties of questions that succeeded, Ponzanelli et~al.~\cite{ponzanelli2014understanding} classify question quality from textual and community features, and Bhat et~al.~\cite{bhat2014min} relate tag choice to response time; \Cref{sec:rq4} measures the same kinds of properties against the asker's standing on the same questions and finds that the standing dominates. del Rio-Chanona et~al.~\cite{del2024large} measure the fall in Stack Overflow activity after the release of ChatGPT against comparison platforms; \Cref{sec:disc_decay} shows that on \ese the closure decline predates such assistants and places the volume collapse beside that finding and the market cycle without separating them.

\section{Conclusion}
\label{sec:conclusion}

This paper reads a decade of revealed developer needs on \ese: \nquestions questions labeled into \nuniquetopics topics at a verified accuracy of 87.9 to 93.5\%. Developers ask, above all, about writing, deploying, and debugging contracts, and toolchain and integration questions form the largest and slowest-answered cluster. As a live-support venue the platform has weakened, and unevenly: answers still arrive, closure falls further, and the decline lies partly in who asks and partly in how each group of askers is served, not in the task mix. Documentation and tooling effort belongs with the toolchain, security tooling with the questions where incident-class mistakes arrive, and newcomer support with context capture and routing; closure was falling before assistants arrived.

\bibliographystyle{IEEEtran}
\bibliography{references}

\end{document}